\documentclass[sigconf]{acmart} 

\AtBeginDocument{%
  \providecommand\BibTeX{{%
    \normalfont B\kern-0.5em{\scshape i\kern-0.25em b}\kern-0.8em\TeX}}}

\setcopyright{acmcopyright}
\copyrightyear{2021}
\acmYear{2021}
\acmDOI{10.1145/1122445.1122456}

\acmConference[Woodstock '18]{Woodstock '18: ACM Symposium on Neural
  Gaze Detection}{June 03--05, 2018}{Woodstock, NY}
\acmBooktitle{Woodstock '18: ACM Symposium on Neural Gaze Detection,
  June 03--05, 2018, Woodstock, NY}
\acmPrice{15.00}
\acmISBN{978-1-4503-XXXX-X/18/06}

\usepackage{multirow}

\usepackage{pgfplots}
\pgfplotsset{width=9cm,compat=1.9}

\usepgfplotslibrary{external}
\usepackage{tabularx}
\usepackage{enumitem}

\newcolumntype{b}{X}
\newcolumntype{s}{>{\hsize=.25\hsize}X}
\newcolumntype{m}{>{\hsize=.5\hsize}X}
\begin{document}

\title{Profiler: Profile-Based Model to Detect Phishing Emails}

\author{Mariya Shmalko}
\email{m.shmalko@unsw.edu.au}
\affiliation{%
  \institution{University of New South Wales  Cybersecurity CRC}
}
\author{Alsharif Abuadbba}
\email{sharif.abuadbba@data61.csiro.au}
\affiliation{%
  \institution{CSIRO's Data61}
}
\author{Raj Gaire}
\email{raj.gaire@data61.csiro.au}
\affiliation{%
  \institution{CSIRO's Data61}
}
\author{Tingmin Wu}
\email{tina.wu@data61.csiro.au}
\affiliation{%
  \institution{CSIRO's Data61}
}
\author{Hye-Young Paik}
\email{h.paik@unsw.edu.au}
\affiliation{%
  \institution{University of New South Wales}
}
\author{Surya Nepal}
\email{surya.nepal@data61.csiro.au}
\affiliation{%
  \institution{CSIRO's Data61 \& Cybersecurity CRC}
}

\renewcommand{\shortauthors}{Trovato and Tobin, et al.}
\newcommand{\sharif}[1]{\textcolor{blue}{[Sharif: #1]}}
\newcommand{\helen}[1]{\textcolor{blue}{[Helen: #1]}}
\newcommand{\tina}[1]{\textcolor{teal}{[Tina: #1]}}
\setcopyright{none}
\settopmatter{printacmref=false} 
\renewcommand\footnotetextcopyrightpermission[1]{} 
\pagestyle{plain}

\begin{abstract}

Due to the popularity of email as a form of communication, email phishing has become more prevalent and grows more sophisticated over time. To combat this rise, many machine learning algorithms for detecting phishing emails have been developed and deployed in the real world. However, due to the limited email data sets on which these algorithms train, they are not adept at recognising varied attacks and, thus, suffer from concept drift; attackers can introduce small changes in the statistical characteristics of their emails or websites to successfully bypass detection. Over time, a gap develops between the reported accuracy from literature and the algorithm's actual effectiveness in the real world \cite{realwordVsReported}. This realises itself in frequent false positive and false negative classifications.

To this end, we propose a multidimensional risk assessment of emails to reduce the feasibility of an attacker adapting their email and avoiding detection. This horizontal approach to email phishing detection profiles an incoming email on its main features. We develop a risk assessment framework that includes three models which analyse an email's (1) threat level, (2) cognitive manipulation, and (3) email type, which we combine to return the final risk assessment score. The Profiler does not require large data sets to train on to be effective and its analysis of varied email features reduces the impact of concept drift. The Profiler's output can be used as a detection system to flag emails that are assessed as high risk at an early stage. Our Profiler can also be used in conjunction with machine learning approaches, to reduce their misclassifications or as a labeller for large email data sets in the training stage.

We evaluate the efficacy of the Profiler against a machine learning ensemble using state-of-the-art machine learning algorithms on a data set of 9000 legitimate and 900 phishing emails from a large Australian research organisation. Our results indicate that the Profiler's horizontal approach delivers 30$\%$ less false positive and 25\% less false negative email classifications over the machine learning ensemble's approach. We show that the Profiler mitigates the impact of concept drift by applying it to a new form of phishing. To do this, we use a selected data set of 3300 file sharing phishing emails which we run against our machine learning ensemble baseline and the Profiler. We also discuss the implications of our results for future email phishing detection. 
\end{abstract}

\keywords{Email Phishing, Concept Drift, Orchestration, Profiling}

\maketitle

\section{Introduction}\label{sec:intro} 


Email phishing is the most prevalent form of phishing attack which exists today \cite{Tessian}. Under the guise of a trustworthy entity, attackers manipulate human factors to steal sensitive information or install malware on a target user's device. The underlying goal is often financially motivating \cite{verizon} and can cost businesses up to \$5.81 million AUD \cite{IBM2021Report}. Taking into consideration COVID-19 and its impact on business work practices, the average cost of a data breach is \$1.47 million AUD higher in breaches where remote work is a factor \cite{IBM2021Report}. Evidently, there is great motivation to find methods which can detect email phishing and curb the risks associated with it. 


There are a wide range of existing technical solutions to email phishing which generally fall under two categories: heuristic approaches and machine learning~\cite{abuadbba2022towards}. Heuristic approaches leverage known email phishing formats to develop rules which can capture phishing emails. These solutions can provide blanket coverage, however, have been shown to be less effective against more sophisticated attacks \cite{blacklistsUneffective,almashor2021characterizing}. Thus, machine learning solutions were developed to target certain forms of email phishing or specific aspects of the email phishing flow. Some examples of existing machine learning methods include natural language processing on the text within an email \cite{fang2019themis, language-spear,evans2021raider,kashapov2022email}, URL analysis \cite{le2018urlnet}, HTML \cite{opara2019htmlphish, HTMLstacking}, and visual analysis \cite{phishpedia} of web pages which could be linked via email.  


A majority of works cite machine learning as the go-to phishing detection solution because of its ability to continuously learn and we see large email services providers preferencing machine learning over heuristic approaches (details in Section \ref{section:emailServices}). 

However, we highlight two challenges in using machine learning approaches alone.
First, these machine learning models need large collections of email data sets, both legitimate and phishing emails, to train on. Given the sensitive nature of emails, gaining access to these data sets is very challenging. Second, due to the evolving nature of emails, machine learning models suffer from concept drift and thus, over time, become less effective at detecting email phishing. We discuss these problems in more detail in Section~\ref{sec:bg:ml}.

%
%
To solve the lack of email data on which to train on and reduce the effect of concept drifting, we propose a horizontal approach to email phishing detection - heuristically profiling an email on its features to determine the risk of it being a phishing attack. By performing a multidimensional risk assessment of an incoming email, we reduce the feasibility of an attacker being able to deceive our Profiler on multiple grounds and, thus, make our solution more resilient to concept drift and adaptive attacks.  

Our Profiler has three models which analyse an email on its (1) threat level, (2) cognitive manipulation, and (3) email type, returning a positive numerical output. We have selected these three models, to demonstrate the benefits of a horizontal phishing detection approach, as they cover the main aspects of an email: its body, subject, sender and receiver. The analysis of focused data sets that can capture phishing email features is crucial in this approach. As such, there is no need for training on data sets and the Profiler does not require large volumes of data in order to be successful. Further, the Profiler does not do a vertical deep dive into each email feature. With a modular design, it can easily target new forms of email phishing and reduce its susceptibility to concept drifting. The Profiler's numerical output creates an opportunity for a sliding threshold that can be used to triage emails and can be adjusted to adapt to the constant evolution of phishing attacks and, thus, reduce the number of false positive and false negative classifications.


In order to validate the usefulness of our approach, we selected and trained three state-of-the-art machine learning models in an ensemble (details in Section \ref{section:ensemble}) to use as a baseline against the Profiler: THEMIS \cite{fang2019themis}, URLNet \cite{le2018urlnet} and HTMLPhish \cite{opara2019htmlphish}. We compared the results of our Profiler to the ensemble on three real-time data sets collected from a large Australian research organisation, including: (1) 9000 legitimate emails, (2) 900 reported phishing emails, and (3) 3300 reported file sharing phishing emails. Our study found that the Profiler has a 17\% false-positive rate, reduced from the ensemble's 47\% and a 25\% false-negative rate, reduced from the ensemble's 45\%.

The Profiler framework can be complementary to existing machine learning models to improve email phishing detection. For instance, the Profiler could be used as a reliable triaging mechanism to filter out false positive and false negative emails. Furthermore, the Profiler can be used in the training stage of machine learning models as an automatic labeller of training data in order to reduce time spent on training new machine learning models.

To this end, our paper makes the following contributions: 
\begin{itemize}
    \item We investigate the behavior of a combination of state-of-the-art machine learning models which look at various dimensions of emails and how these perform in real time. We identify the problem of concept drift in email phishing detection; there is a vast difference between training-time accuracy and real-time accuracy.
    \item We develop the first orchestrator framework which heuristically profiles an email on three components: (1) threat level, (2) cognitive manipulation, and (3) email type. Our framework uses a score based mechanism to decide the legitimacy of the email. It identifies ``signals'' through email profiling rather than pre-training to reduce the impact of a lack of email data. The Profiler's modular design also tackles the issue of concept drift. 
    \item We propose the Profiler could be used as a reliable triaging mechanism to filter out false positive and false negative classifications before applying further machine learning solutions to email inputs. Or, as a labeller of email data sets to help in the training stage when developing new machine learning models. 
    \item We perform a thorough evaluation following two use case studies: (a) a collection of a mixture of 9000 legitimate and 900 phishing emails collected from a large Australian research organisation, and (b) a private data set containing 3300 file sharing phishing emails - an emerging form of phishing. Our results demonstrate that we could achieve a 30\% decrease in false positive classifications and a 25\% decrease in false negative classifications as when compared to existing machine learning solutions.
\end{itemize}

Section \ref{section:background} introduces the background and related work. Section \ref{section:approach} will present the Profiler overview and Section \ref{section:impl} will discuss implementation details. Section \ref{section:evaluation} will explain the experimental setup and Section \ref{section:results} will present the results. Finally, Section \ref{section:discussion} will discuss the benefits of the Profiler and present future work.
    
\section{Background and Related Work}
\label{section:background}
We start by introducing email phishing, and existing solutions to combat it. We then present the solutions taken by leading email services. We also discuss closely related works relevant to this paper.

\subsection{Email Phishing}
Email is a common form of communication that has been exploited as a method for phishing attacks \cite{FBI2020CrimeReport}, with more than 11 times the number of phishing complaints in 2020 compared to just 4 years prior in 2016 \cite{FBI2020CrimeReport}. Attackers disguise their email address, posing as a trustworthy entity and attempt to manipulate the victim to steal sensitive information, for financial gain or to deploy malicious software onto the victim's infrastructure \cite{verizon}. 

Given it is a social attack, and inbox owners are becoming increasingly aware of the possibility of a phishing email, email phishing attack vectors are constantly evolving and have become increasingly sophisticated over time. Attacks have advanced from offering monetary rewards as an incentive for the victim to reply, to emails containing hidden links to legitimate-looking websites which are under the control of the attacker. These websites lure the victim into entering personal or financial information which can be stored and used against them. 

Email phishing extends further than stealing an individuals information. Phishers are often targeting the employees of large organisations to obtain authentication credentials to perform a lateral phishing attack \cite{lateralPhishing}. Once credentials are acquired, phishers can carry out malicious attacks against the organisation, rather than the employee. This results in expensive data breaches and can lead to the leaking of customer data, intellectual property and internal documents costing the organisation millions \cite{IBM2021Report}.

There is motivation to find effective email phishing prevention methods which would protect against such attacks. A wide range of techniques has been developed to prevent phishing attacks from impacting users which generally fall into two categories: heuristic approaches which take advantage of known email phishing attack forms and machine learning models which are trained on existing email phishing data sets. 

\subsection{Heuristic Solutions}

Heuristic approaches study the features of known phishing emails to design models which can detect phishing on the basis of these features. A common heuristic approach includes maintaining a blacklist of known phishing URLs or domains and checking the sender and URLs of a received email against these lists \cite{blacklist}. Rules can be created to target all aspects of the email phishing flow. Approaches may analyse URL patterns \cite{blacklistURLs} \cite{heuristicURLs}or perform a filtering of words using a bag-of-words algorithm \cite{bagofwords}. Malware analysis is also an area where a heuristic approach may be appropriate \cite{googlemalware}. 

Well designed heuristic approaches are able to accept more rules to ensure an advanced analysis of the attacks and to better modify the detection of email phishing \cite{vade}. These approaches leverage the knowledge of existing phishing attacks and are capable of providing solid initial coverage from low-effort email phishing attempts \cite{anatomyOfPhishing}. However, they have been shown to not be as effective against more sophisticated attacks \cite{blacklistsUneffective}. 

This is where machine learning approaches are used. Targeting certain forms of email phishing and specific aspects of the email phishing flow, machine learning models are able to determine a binary classification: phishing or legitimate.

\subsection{Machine Learning Solutions}
\label{sec:bg:ml}
Existing machine learning solutions follow a vertical approach to phishing detection - focusing specifically on a single feature within an email. Models look at the email's text\cite{fang2019themis} \cite{language-spear}, URLs \cite{le2018urlnet}, HTML structure of linked pages\cite{opara2019htmlphish} \cite{HTMLstacking}, or visual appearance of these pages \cite{phishpedia}. These solutions learn from experience and train using large labelled data sets of both legitimate and phishing content so that models can recognise future unlabelled data \cite{mldef}. 

Machine learning models which learn to identify phishing through the email's text or subject split the email into words or characters to identify spelling inconsistencies, general email structure or phrasing \cite{fang2019themis}. They select words that are determined to be highly valuable to make a classification. Similarly, models which aim to capture phishing URLs, may split the URL into characters or words to analyse key findings \cite{le2018urlnet}. HTML analysis models inspect HTML tags \cite{opara2019htmlphish} or content within tags while visual analysis models use markers to identify key features of the web page or logo being examined \cite{phishpedia}. 


While these methods achieve high reported success, we highlight some of the challenges with existing machine learning-based solutions. 

First, machine learning models require large, feature-engineered data sets, which include legitimate and phishing emails, to train on. Ideally, these data sets would capture the current state of email phishing. This is to ensure that the models are effective on unseen data when applied in the real world and are able to recognise current trends. Given the sensitive nature of emails and strict privacy rules such as the GDPR and the Australian Privacy Act, it can be challenging to gain access to these data sets. Individuals' work email addresses are classified as personal data, given they usually contain an identifying portion of the employee's name, and, as such, are not often released for research purposes \cite{gdpr}. Further, organisations are hesitant to volunteer their emails to third-party researchers as they can contain private information which cannot be revealed. As such, access to large and varied data sets which capture the current state of email phishing attacks are not readily available making training of machine learning models a difficult process. 


Second, the rapid evolution that phishing emails go through to bypass detection methods results in machine learning models suffering from concept drift. A requirement for building successful machine learning models is the ability to train on a wide variety of samples such that you can generalise the model to adapt to unseen, future behaviours \cite{ConceptDrift}. However, phishing evolves rapidly and attackers are able to effectively test their attacks before deploying them. For example, there are websites that will check if an email could bypass existing security measures or would be flagged by several leading email services such as Gmail, Outlook and Yahoo \cite{spamChecker}. Thus, as attackers are able to easily modify their emails to avoid detection, it is difficult to establish models such that they are efficient on unseen behaviours. This issue is compounded with the lack of available data to train on. As such, over time, machine learning models which trained vertically to recognise certain aspects of phishing emails may become obsolete as attackers defer from known methods. 

Moreover, a crucial requirement for building successful machine learning models is a large variety of data. While we've already discussed the difficulty of gaining access to email data sets, if access \textit{were} to be granted, there are not enough samples of each new form of email phishing to accurately capture the method and re-train models to recognise it. Essentially, machine learning models are `playing catch-up' with new forms of email phishing. 


Finally, some machine learning models can be circumvented altogether. As an example, consider a model which analyses the HTML of a web page but the web page is built using Javascript to specifically obfuscate the web page's purpose \cite{proofpoint}. In this case, the HTML analysis would not return any valuable results. Once attackers are aware that the HTML of their phishing websites is being scanned, they can start building Javascript websites and the HTML analysis model becomes obsolete.  

Taking another example, a machine learning model which analyses the visual appearance of a web page; it can only protect email receivers from phishing attacks that attempt to mimic \textit{top} websites - those that are most commonly used for phishing, such as LinkedIn, Amazon or PayPal \cite{phishpedia}. Other, less well-known brands can still be impersonated without detection. 

Thus, while machine learning solutions demonstrate robustness on offline data sets which have been cleaned, feature engineered and focused onto a specific aspect of the email phishing attack flow, they suffer in real-time scenarios due to a lack of varied data and the constant evolution of email phishing attack vectors which leads to concept drift. This is realised in frequent false positive and false negative classifications and has been demonstrated in our own work in Section \ref{section:results}.

\subsection{Leading Email Services and Email Phishing}
\label{section:emailServices}
Given the high cost and prevalence of email phishing, large email services provide protection against phishing in the form of banner warnings or automatic filtering of emails. However, the implementation of their models is a black box and information on how email phishing detection is performed by these services is not readily available. The most popular email service in 2020 was Gmail \cite{popularEmailServices}, with around 1.8 billion users \cite{gmailUsers}. The second most popular service was NetEase Mail which is the product of a Chinese internet company and the third most common was Microsoft's Outlook \cite{popularEmailServices}. Given that Gmail and Outlook are most familiar to English speakers, below we discuss the available information on how Google and Microsoft detect and respond to email phishing attacks. This information is valuable as it demonstrates reasons why commercial products perform at a higher rate of accuracy than those reported in literature. 

\subsubsection{Gmail}
Google uses machine learning to block spam and phishing emails from appearing in a user's inbox with an accuracy of 99.9\% \cite{google2017protectagainstphishing}. They use early phishing detection, a dedicated machine learning model, to analyse each email received and check for malicious intent. This process delays an email appearing in the inbox by 0.5 seconds \cite{google2017delayeddelivery}. The phishing detection model is also integrated with Google's Safe Browsing machine learning technology which detects suspicious websites, and hence, their URLs. The early phishing detection compares any URL received within an email with those which have been previously flagged by Google as suspicious. They also check for malicious attachments and identify unauthenticated emails \cite{google2019attachments}.  
Given Google's unlimited access to emails and websites, their machine learning algorithms have a wide variety of data to train on. As such, we speculate that their models do not suffer from concept drift and thus have a lower rate of false positive and false negative classifications. Further, given their extensive architecture, they can run their model on each email with minimal impact on performance and can afford to retrain their models efficiently and on varied data. 

\subsubsection{Outlook}
Like Google, Microsoft also has a phishing detection mechanism that they run on their Outlook product. They use machine learning models to detect malicious attachments and query URLs in real-time so that malicious ones may be detected \cite{microsoft2018machinelearning}. Outlook will also warn users of suspicious domain names in email addresses \cite{microsoft2018suspmessages}. 

By default, Outlook sets the inbox filtering to `low' so it will move only the ``most obvious'' junk mail into the junk email folder. The `high' level of filtering is described by Microsoft as, ``Most junk email is caught, but some regular mail may be caught as well.'' Microsoft's machine learning model is not as sophisticated as Google's may be and produces false positive classifications \cite{microsoft2018levels}. However, Microsoft still benefits from the large amount of email data that they can use to train their models to reduce the impact of concept drift. 

\subsection{Existing Profiling}
In this section we discuss two related works which are relevant to this paper: Cognitive Triaging of Phishing Attacks \cite{van2019cognitive} and Categorizing Human Phishing Difficulty: A Phish Scale \cite{phishscale2020steves} .

\subsubsection{Cognitive Triaging of Phishing Emails}
\label{sec:cogbackground}
In the context of this work, cognitive triaging is used in the Cognitive Assessment Model of the Profiler, explained in more detail in Section \ref{sec:cognitiveAssessment}. 

Cognitive triaging refers to a quantitative measure of cognitive vulnerability triggers which come from psychology literature. There are six principles of influence: scarcity, consistency, social proof, authority, liking and reciprocity which can impact the likelihood of a human's decision on whether or not to perform the action they were requested to do \cite{cialdini}.

The paper predicted the degree of success of a phishing attack using these vulnerability triggers, measured by how likely employees were to interact with emails and found that emails that exploited the scarcity and consistency vulnerabilities were often deemed as legitimate by users and had a high level of interaction.

Building on the findings of this paper, we adopt the scarcity and consistency vulnerabilities into our Cognitive Assessment model. The reported measure on the vulnerability triggers helped determine the vulnerabilities which we chose to explore in our risk-based model to further analyse the manipulative language within an email. 

\subsubsection{A Phish Scale}
\label{sec:phishscalebackground}
Phish Scale was developed as a tool to easily categorise the difficulty of recognising a phishing email \cite{phishscale2020steves}. It looks at two components in order to develop this rating: the cues of the email and the alignment of the email's context to the user. An email with fewer cues and a more relevant context would be classified as `high' on the phish scale and hence more difficult to detect as phishing.

Phish Scale does not do email phishing detection, rather they categorise known phishing attacks into `low', `medium' or `high' groups based on their two components. In contrast, in our work we produce a numerical risk assessment score for an email determined from three models which focus on profiling an email.

\section{Profiler System Design}
\label{section:approach}

To recap, existing academic works to solve email phishing cite machine learning as the preferred solution to email phishing detection. Many models have been and are being developed to work vertically - looking into a specific feature of the email such as URLs, HTML or language. These works often have small data sets which cannot capture the different forms of email phishing and over time, as phishers adapt, these models begin to suffer from concept drift and their accuracy falls. 

To solve the lack of data on which to train on and to lower the effect of concept drift, we explore a horizontal approach to email phishing detection; involving three heuristic models that profile an incoming email on its features. Our Profiler generates a numerical risk assessment score from analysing the (1) \textit{threat level} (details in Section \ref{sec:threat_level}), (2) \textit{cognitive manipulation} (details in Section \ref{sec:cognitiveAssessment}) and (3) \textit{email type} (details in Section \ref{sec:emailProfile}). Our approach does not require large data sets to train on, rather we analyse focused data sets to capture phishing email features. Further, the models chosen in this work cover a multitude of features within an email: its subject, body, sender and receiver. As we do not do a vertical deep dive into each of these, instead choosing to perform a multidimensional risk assessment on many aspects of an email, our method is able to diminish the impact of rapidly evolving forms of phishing and can easily adapt to capture them. Thus, this method is less susceptible to concept drift than existing machine learning solutions. Finally, the numerical output of our Profiler, can be used to triage the email and determine its risk - the likelihood of phishing in order to reduce false positive and false negative classifications. Or, the Profiler can be used to automatically label data sets to help in the training stage of machine learning approaches. 

\begin{figure}[H]
    \centering
    \includegraphics[width=0.45\textwidth,scale=1.1]{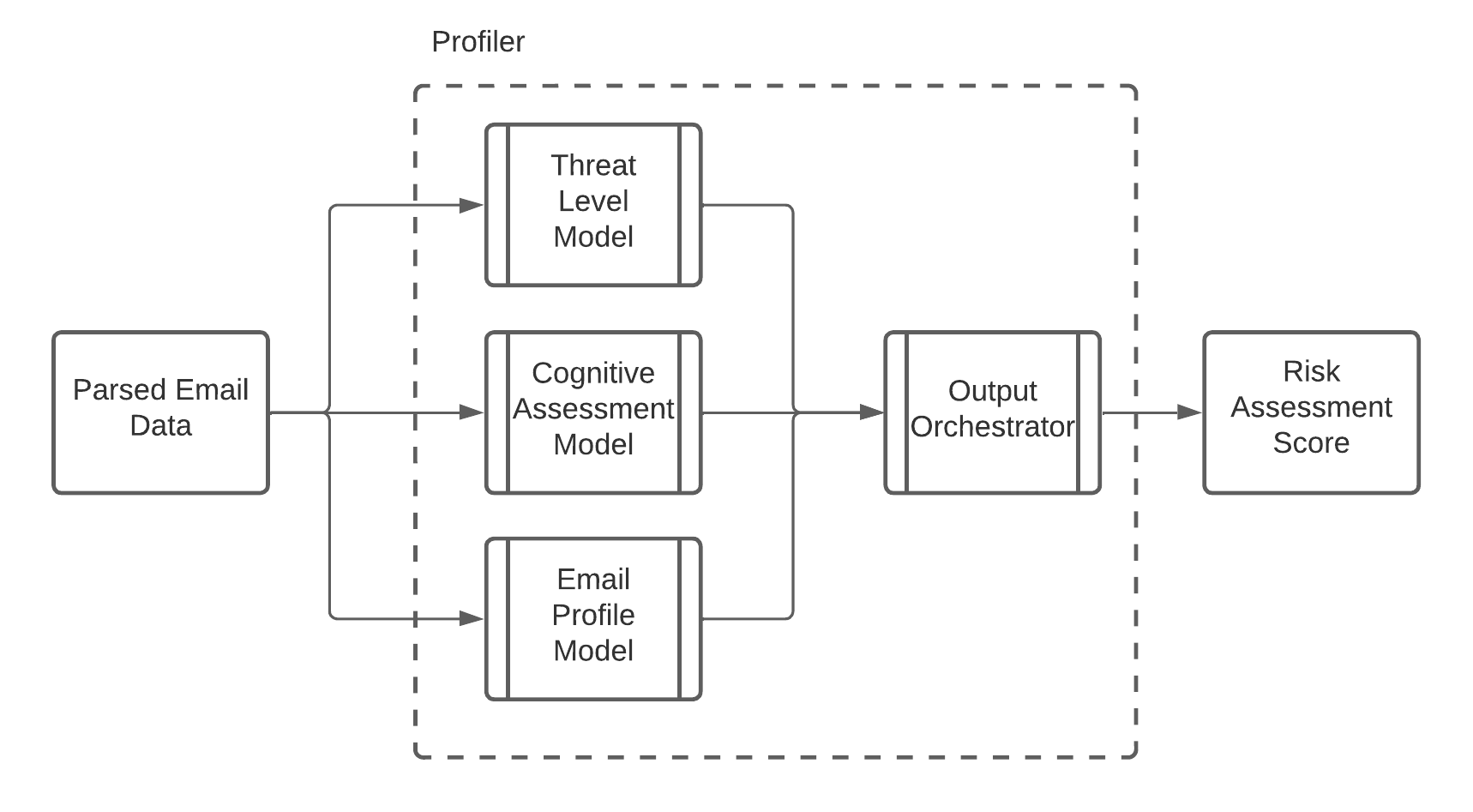}
    \caption{Profiler Overview.}
    \label{fig:profiler_overview}
\end{figure}

\textbf{Overview.} An overview of the Profiler is shown in Figure \ref{fig:profiler_overview}. An email's information is parsed and decoded and then passed to each of the three models. Each model performs its own calculations to determine the risk of an email from its perspective and returns a numerical output. Each model's output is passed to the output orchestrator which produces the final risk assessment score that can then be used to classify an email. 

There are three models, which assess an email's sender, receiver, subject and body. Each model was developed to target a specific aspect of the email to ensure a multidimensional risk assessment. We have selected these three models to demonstrate the benefits of a horizontal phishing detection approach, as they cover the main aspects of an email. We now introduce the details of each model. 


\subsection{Threat Level Model}\label{sec:threat_level}

An email address has the structure: \textit{username}@\textit{domain} and can be split into two parts. Our threat level model looks at both the username and domain of an email address to heuristically determine malicious intent. Its purpose is to establish the source of the email and its threat level. 

The model itself returns two values: `low' and `high'. A `low' score corresponds to the numerical value of 0.1 and a `high' score corresponds to the numerical value of 0.9. We chose two values between 0 and 1 for simplicity. 

\begin{figure}[H]
    \centering
    \includegraphics[width=8.5cm]{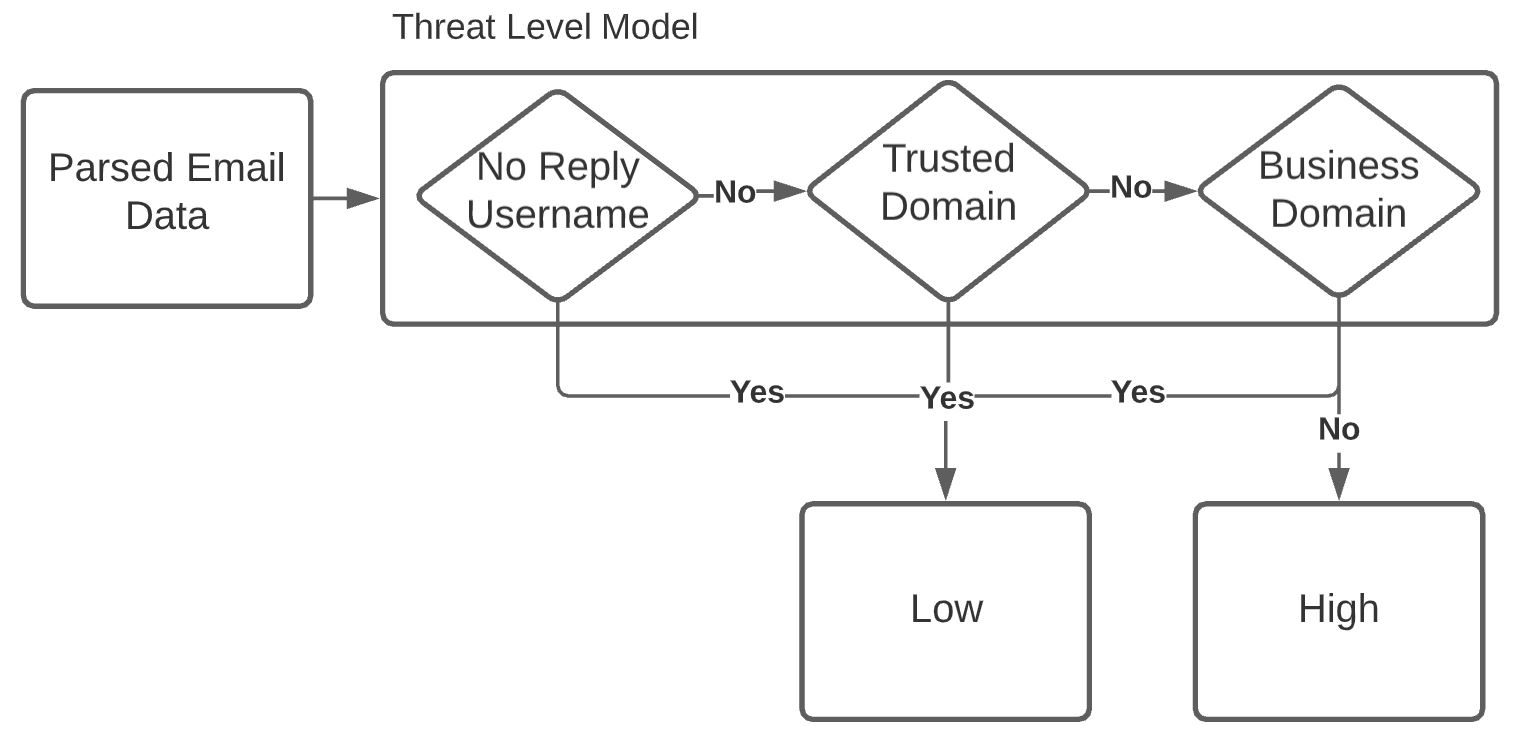}
    \caption{Threat Level Model Overview.}
    \label{fig:trustassessment}
\end{figure}

We look at three features of the sender's username and domain: no reply username, trusted domain and business domain. The model flow can be seen in Figure \ref{fig:trustassessment}. The definitions of each feature can be seen in Table \ref{table:trustassessment}.

\begin{table}
    \begin{center}
    \caption{Trust Assessment Features Summary}
    \label{table:trustassessment}
    \begin{tabularx}{.5\textwidth}{lX}
        \hline 
        \textbf{Feature} & \textbf{Definition} \\
        \hline
        \textbf{No Reply Username}  & Username contains any variation of ’no-reply’. Ignoring all white space, hyphens and capitalisation. \\
        \hline
        \textbf{Trusted Domain} & Any of: \begin{itemize}[leftmargin=*]
            \item \textbf{Internal Domain} \newline Domain is the same as receiving inbox domain
            \item \textbf{Australian Government Domain} \newline Domain contains the key phrase 'gov.au'
            \item \textbf{Australian Education Domain} \newline Domain contains the key phrase 'edu.au'
        \end{itemize} \\
        \hline
        \textbf{Business Domain} & Domain contains the key phrase ’.com’ or ’.org’  
    \end{tabularx}
    \end{center}
\end{table}

As the emails from our data sets were Australian based, we chose to focus on Australia-specific email domains. However, this could be easily generalised to a global scale.

\subsubsection{Model Output}
The model will return a `low' value of 0.1 if the sender username is a no-reply username, or if the sender domain is an internal domain or a trusted domain. Otherwise, the model will return a `high' value of 0.9, indicating that there is a higher risk of threat associated with this sender.

\subsection{Cognitive Assessment Model}\label{sec:cognitiveAssessment}

The purpose of the cognitive assessment model is to analyse the email's subject and body to assess whether or not the sender is trying to make use of human psychological vulnerabilities to manipulate the email receiver. 

This model consists of two parts: the cognitive triaging analysis and the monetary incentives analysis.

\subsubsection{Cognitive Triaging} 
This part of the model is based on findings from the paper `Cognitive Triaging of Phishing Attacks' discussed in Section \ref{section:background} of this paper. Given that the paper only found a high correlation between consistency and scarcity, these are the two categories of words on which this model concentrates.

Scarcity refers to the limiting of resources, for example ``\textit{Your account will be shut down immediately if you do not reply within 24 hours.}" Consistency refers to honouring previous agreements. An email containing the line ``\textit{You have previously agreed to terms and conditions, please continue to follow them. Click here to unflag your account}," would fall under this category of cognitive manipulation. 

This assessment is a bag-of-words approach that analyses the frequency of key words, and returns the count as the score given by this model. A higher frequency indicates an email that is using more persuasive language and, thus, should be treated as more suspicious. A lower frequency indicates lower risk - cognitive manipulation is not prevalent.

\subsubsection{Monetary Incentives}
An `unexpected money' scam offers its victim the false promise of a monetary reward in order to trick them into sharing banking details or parting with money. In Australia, \$3 million AUD was lost to this kind of scam in 2020 alone \cite{unexpectedmoney} and an email is a convenient form that helps scammers target their victims. As such, we have implemented a monetary incentives analysis. This is also a frequency analysis of key words present in the email subject and body specific to monetary incentives. Examples of key words include `millionaire', `reward' or `payment'.

\subsubsection{Model Output}
Our Cognitive Assessment model returns the sum of the Cognitive Triaging word count and Monetary Incentives word count. Emails which have a higher count are treated as more suspicious. 

\subsection{Email Profile Assessment Model}\label{sec:emailProfile}

The Email Profile Assessment model's purpose is to gauge the context in which the email is sent. This model looks for key words within the subject of an email and applies a simple label to it. Each label has a different weighting which then impacts the final scoring. 

This labelling can help contextualise the email and determine its purpose, and thus, its risk. An email that is a simple receipt for an online order is less likely to be phishing than one which has a congratulatory tone. Further, the likelihood a receiver would interact with an email is considered to refine the output of this model.

The different profile types, their definitions and the score which they return can be seen in Table \ref{table:profiles}.
\begin{table}
    \begin{center}
    \caption{Email Profiles Summary}
    \label{table:profiles}
    \begin{tabularx}{.5\textwidth}{msb}
        \hline
        \textbf{Email Profile} & \textbf{Score} & \textbf{Definition} \\
        \hline
        \textbf{Welcome} & 0.1 & An email which is received after joining an online service or platform \\
        \hline
        \textbf{Work} & 0.1 & An email which notifies of meetings or newsletters \\
        \hline
        \textbf{Job Search} & 0.3 & An email which is sent containing job searches or job related content \\
        \hline
        \textbf{Update} & 0.5 & An email which is sent to nudge or remind the user \\
        \hline
        \textbf{Receipt} & 0.9 & An email which is received after a purchase, usually contains the itemised list and total price paid \\
        \hline
        \textbf{Congratulatory} & 2 & An email which is sent in congratulations, usually this form of email is spam or phishing \\
        \hline
        \textbf{Delivery} & 2 & An email which is sent to notify of an upcoming or past delivery \\
        \hline
        \textbf{Other} & 1 & All other emails
    \end{tabularx}
    \end{center}
\end{table}

\subsubsection{Model Output}
The model returns its associated weighting for the labelled email profile type.

This model is quite flexible and can be adjusted easily if needed. The values associated with each category were selected on a scale between 0 and 5. Given the output of our Profiler is the product of each model's output, the default value for our email profiling model must be 1. All other values were selected in reference to this. 

\subsection{Output Orchestration}
The Profiler takes the product of all scores produced from the three models in order to obtain the final risk assessment score. This is done for simplicity and permits for further extension of the Profiler if more models were to be added. It also allows the application of weightings to impact the influence a model has on the final output. The benefits of this adaptability and modularity are discussed more in Section \ref{section:discussion}.

\section{Implementation}
\label{section:impl}
In this section we discuss the technical implementation and challenges which arise when working with email data. 

For this work, we implemented both our Profiler and our machine learning ensemble in Python 3.7.0. In order to utilise our email data sets effectively, we had to extract the relevant information, decode and pre-process it before passing it to our models. The challenges are discussed below.

\subsection{Parsing Email Data}
There are several common email file formats that exist to store email data. This work is able to parse \textit{mbox}, \textit{txt} and \textit{pkl} formatted files. However, anyone can leverage the use of the Profiler given they are able to extract the email subject, email body, sender email address and receiver email address in decoded string format from the email file format in which they are storing their data. 

We give a brief description of the \textit{mbox} file format. \textit{mbox} is a format for storing collections of emails in a single file, concatenated one after another. The main benefit of this form of email data storage is the ability to extract features easily. In Python, we use the mailbox module \footnote{https://docs.python.org/3/library/mailbox.html} to access and manipulate the emails contained within an \textit{mbox} file. This format also allows us to easily iterate through our data sets.

\subsection{Decoding Email Text}
Due to the sensitive nature of their content, data such as the email header - which contains metadata about the email, including the sender email address, receiver email address and subject, or the email body may be encoded. For the Profiler to be successful, this data must be decoded. 

In this work, two out of our three data sets required decoding with one having only its email header encoded, while the other had both the email header and email body encoded. 

The emails had a content transfer encoding MIME header file. This type of encoding was developed in 1992 to represent binary data in formats other than ASCII text \cite{contentencoding}. In this case, the header was encoded using base64 which encodes text in arbitrary, non-humanely readable octet sequences. To decode the headers, we made use of the legacy email API module, \textit{email.header} \footnote{https://docs.python.org/3/library/email.header.html}. This module is able to decode the header and output a list of tuples containing the decoded string and the encoding used. Since we only needed the subject, sender, and receiver, this was easily extracted using the \textit{$decode\_header()$} function from this module.

Our second data set was a phishing data set and also had a content transfer encoding MIME header file for each email. Further, each email body was \textit{intentionally} encoded with the same encoding to reduce the risk of interacting with a live phishing link or other potential dangers which may arise when working with phishing emails. 

We decoded it in a similar manner as above - using an inbuilt function. 

\subsection{Pre-Processing Email Text}
After being decoded but before being passed to one of the three models, the email subject and body are run through pre-processing. We first transform every character into lower case and remove all words of length less than or equal to two letters. These words do not provide valuable information and we do not want to take up processing time with them. Further, we remove stop words from both the body and subject of an email. 

Stop words are commonly used words such as \textit{the} or \textit{in}. For our data sets, we use the Python NLTK module \footnote{https://www.nltk.org/} as our list of stop words to filter our email subject and body. Again, these words do not usually provide valuable information, can occur quite frequently, and take up processing time. Removing them doesn't impact the outcome of the Profiler but can result in an increase in performance speed which is why we chose to remove them.

\section{Experimental Set Up and Evaluation}
\label{section:evaluation}

The goal of this work was to demonstrate that a horizontal approach that looks at a variety of aspects of an email is a more appropriate solution than the vertical approach which machine learning models take as there is no need for large data sets and issues of concept drift are mitigated. To do this, we created a machine learning majority-voting ensemble composed of three state-of-the-art machine learning models and ran our available data sets through it. This gave us a baseline level of accuracy for each of the data sets. 

Once the development of the Profiler was complete, we ran it on the same data sets in order to compare its performance against the ensemble's. 

In this section, we discuss the machine learning models used in our baseline ensemble, the threshold chosen for our Profiler and the available data sets on which our experiment was run.

\subsection{Machine Learning Ensemble}
\label{section:ensemble}

Three machine learning algorithms were used in an ensemble with a majority voting in order to establish a baseline for the results. Both the phishing and legitimate data sets were run through this ensemble and if two out of the three algorithms classified the email as phishing, it was labelled as phishing, otherwise it was labelled as legitimate. 

The three algorithms are THEMIS, URLNet and HTMLPhish. The following is a description of each machine learning algorithm utilised.

\subsubsection{THEMIS}
THEMIS uses recurrent convolutional neural networks (RCNN) with multilevel vectors to classify phishing emails with a reported accuracy of 99.848\%. THEMIS looks at an email header and body at a word and character level with the word level focusing on the email content and the character level concentrating on spelling mistakes and personal spelling habits \cite{fang2019themis}.

Re-implementation of THEMIS and training was necessary as its code was not released to the public. We trained our implementation of THEMIS on both phishing and legitimate emails. For our phishing data sets, we used the Nazario phishing corpus \footnote{https://monkey.org/~jose/phishing/}, and staff reported phishing data set containing $38,000$ emails provided by the large Australian research organisation as described in Section \ref{section:datasets}. For the legitimate data set we used the Enron data set \footnote{https://www.cs.cmu.edu/~enron/}. Our implementation achieved results comparable to the \cite{fang2019themis} paper.

\subsubsection{URLNet}
Traditionally, malicious URLs have been detected using blacklists. These blacklists are not exhaustive or robust and cannot detect new and malicious URLs which phishers may employ. As a result, there has been a rise in URL threat assessment using machine learning solutions. One of these is URLNet. 
URLNet applies convolutional neural networks (CNN) to both the characters and words of a URL to learn the URL embedding and detect malicious intent. URLNet has better results than traditional support vector machines (SVM) and existing malicious URL detection methods based on bag-of-words features as it is able to capture the semantic and structural information of the URL \cite{le2018urlnet}. 

The implementation of URLNet is available on GitHub \footnote{https://github.com/Antimalweb/URLNet}. Referring to this implementation, we trained our URLNet on $5,000,000$ phishing URLs collected from VirusTotal \footnote{https://www.virustotal.com} and $5,000,000$ benign URLs retrieved from our internet web crawler.

\subsubsection{HTMLPhish}
HTMLPhish focuses on the HTML analysis of a web page and uses CNNs to learn the semantic dependencies in textual components of the HTML for accurate detection of phishing websites. This model was trained on a data set of 50,000 HTML documents and has around a 93\% reported accuracy, while also being language independent \cite{opara2019htmlphish}. Such an algorithm can be used to evaluate URLs present in emails.

Similar to URLNet, our implementation of HTMLPhish trained on the 5 million phishing websites redirected from URLs collected from VirusTotal and the 5 million benign websites loaded from URLs retrieved from our web crawler. 

\subsubsection{Summary}
All three of these algorithms take a vertical approach to email phishing detection. They focus on a specific aspect of the email and attempt to do a binary classification of it. Further, each model has a reported accuracy of over 90\%, with minimal false positive and false negative classifications. 

\subsection{Profiler Threshold}

In order to make a fair comparison against the machine learning ensemble, we must determine the Profiler's threshold for classification. As our baseline ensemble makes a binary classification, so should the Profiler. Any email with a score below our determined threshold will be labelled as low risk and legitimate, while any email above will be treated as suspicious and labelled as phishing. 

Running emails through the Profiler, we observed most falling within the 0 to 50 range with legitimate emails being clustered around 0 to 0.5 and phishing emails having higher outputs, mainly congregated around a score of 1.

\begin{figure}[h]
    \centering
    \includegraphics[width=1.0\linewidth]{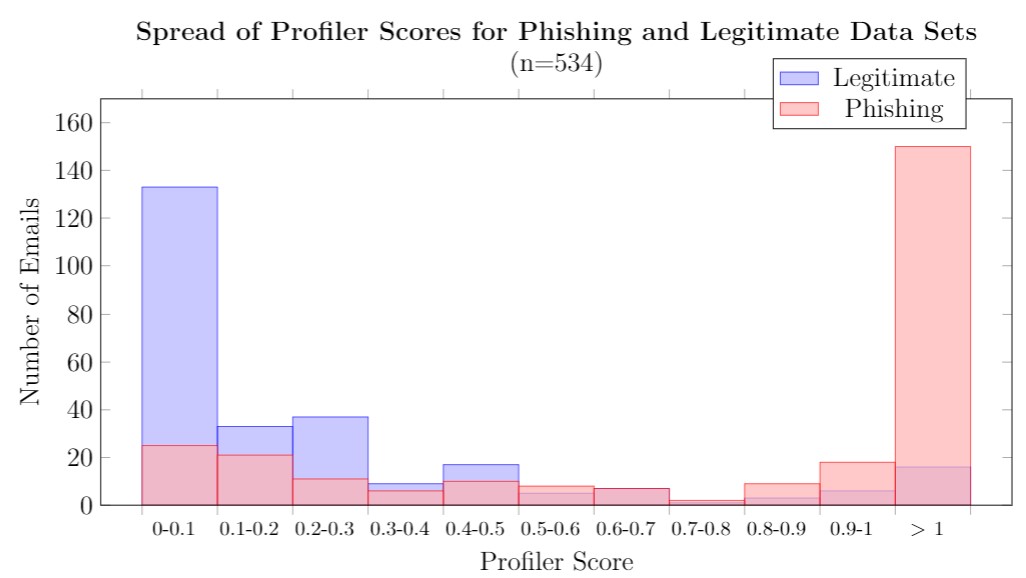}
    \caption{Histogram of Spread of Profiler Scores for Legitimate and Phishing Datasets.}
    \label{fig:histogram}
\end{figure}

To determine the most appropriate threshold, a random 30\% of the phishing data set was taken and matched with an equal quantity of randomly selected emails from the legitimate data set. The scores produced by the Profiler for each email were plotted in a histogram, seen in Figure \ref{fig:histogram}. The histogram is limited in range from 0 to 1 and split into bins of size 0.1. 

Analysing the histogram, we see a steep decrease in legitimate emails and a higher frequency of phishing emails at the $0.5-0.6$ bin. Thus, for the purpose of this work, this is where the threshold was drawn. Any email with a Profiler output of less than 0.5 is legitimate and anything equal to or above is phishing. 

\subsection{Data Sets}
\label{section:datasets}
This work deals with three data sets: legitimate emails, phishing emails and file sharing phishing emails. Table \ref{table:email numbers} details the number of emails in each data set.

\begin{table}
    \begin{center}
    \caption{Number of Emails per Data Set}
    \label{table:email numbers}
    \begin{tabular}{||c|c||} 
        \hline 
        \textbf{Email Data Set} & \textbf{Number of Emails} \\
        \hline \hline
        Legitimate & 9066\\ 
        \hline
        Phishing & 893 \\
        \hline
        File Sharing Phishing & 3292 \\
        \hline
    \end{tabular}
    \end{center}
\end{table}

\subsubsection{Legitimate Data Set}
The legitimate data set contains just over 9000 legitimate emails from various senders. Senders included Facebook, Linkedin, Youtube along with some general personal emails and work-related emails. This data set also contained promotional emails which could be labelled as spam, rather than phishing. 

\subsubsection{Phishing Data Set}
The phishing data set contains just under 1000 emails. These emails are from senders mimicking reputable businesses such as Woolworths, Paypal and Microsoft. The emails in this data set often claimed the money was on offer or subscriptions were expiring. 

\subsubsection{File Sharing Phishing Data Set}
The file sharing phishing data set contains around 3300 emails that were hand labelled and are specific to the file-sharing phishing use case. During the development of the Profiler, this data was not examined to ensure it could be used as an unseen data set for the evaluation of the work.

\section{Results}
\label{section:results}


\subsection{Legitimate Data Set Results}
The legitimate data set contains 9066 legitimate emails. Of these, the machine learning ensemble labeled 52.5\% correctly: 4767 emails were labelled true negative and 4299 were false positives and labelled as phishing.

\begin{figure}[t]
\centering  
\includegraphics[width=0.8\linewidth]{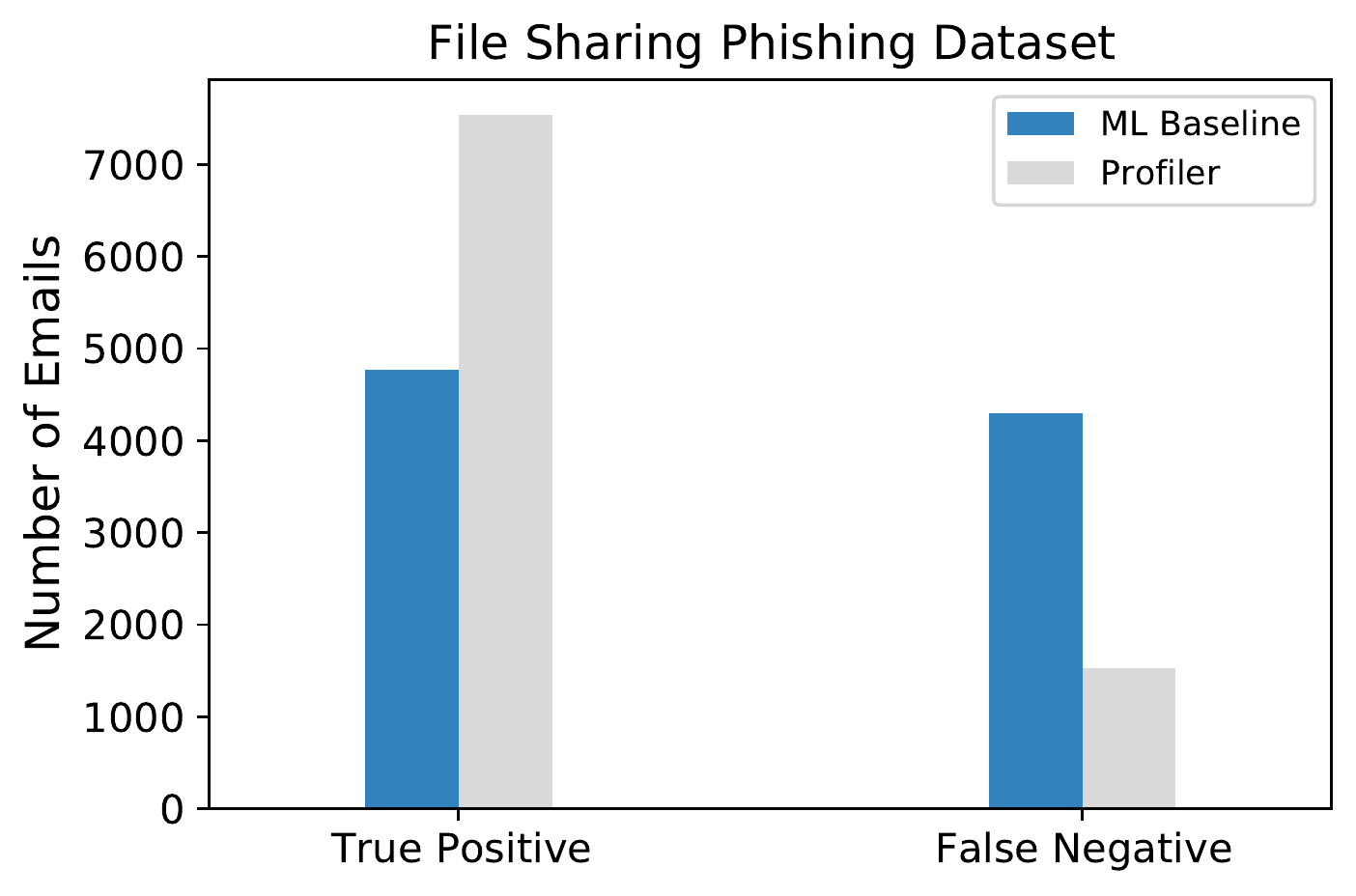}
\caption{Comparing Machine Learning Baseline to Profiler Classifications (n = 9066). True Positive - Phishing labelled Phishing, False Negative - Phishing labelled Legitimate.}
\label{fig:legitresults}
\end{figure}

In contrast, the Profiler labelled 83.2\% of this data set correctly: 7541 labelled true negative and 1525 false positive labels. We can see these results visually in Figure \ref{fig:legitresults} with the true negative column on the left and the false positive column on the right. Evidently, the Profiler is able to classify more emails correctly than the machine learning baseline and achieves a 30\% lower false positive rate. 

\subsection{Phishing Data Set Results}
The phishing data set contains 893 phishing emails. Of these, the machine learning ensemble labeled 69.7\% correctly: 623 emails were labelled true positive and 270 were false negative labels.

\begin{figure}[t]
\centering  
\includegraphics[width=0.8\linewidth]{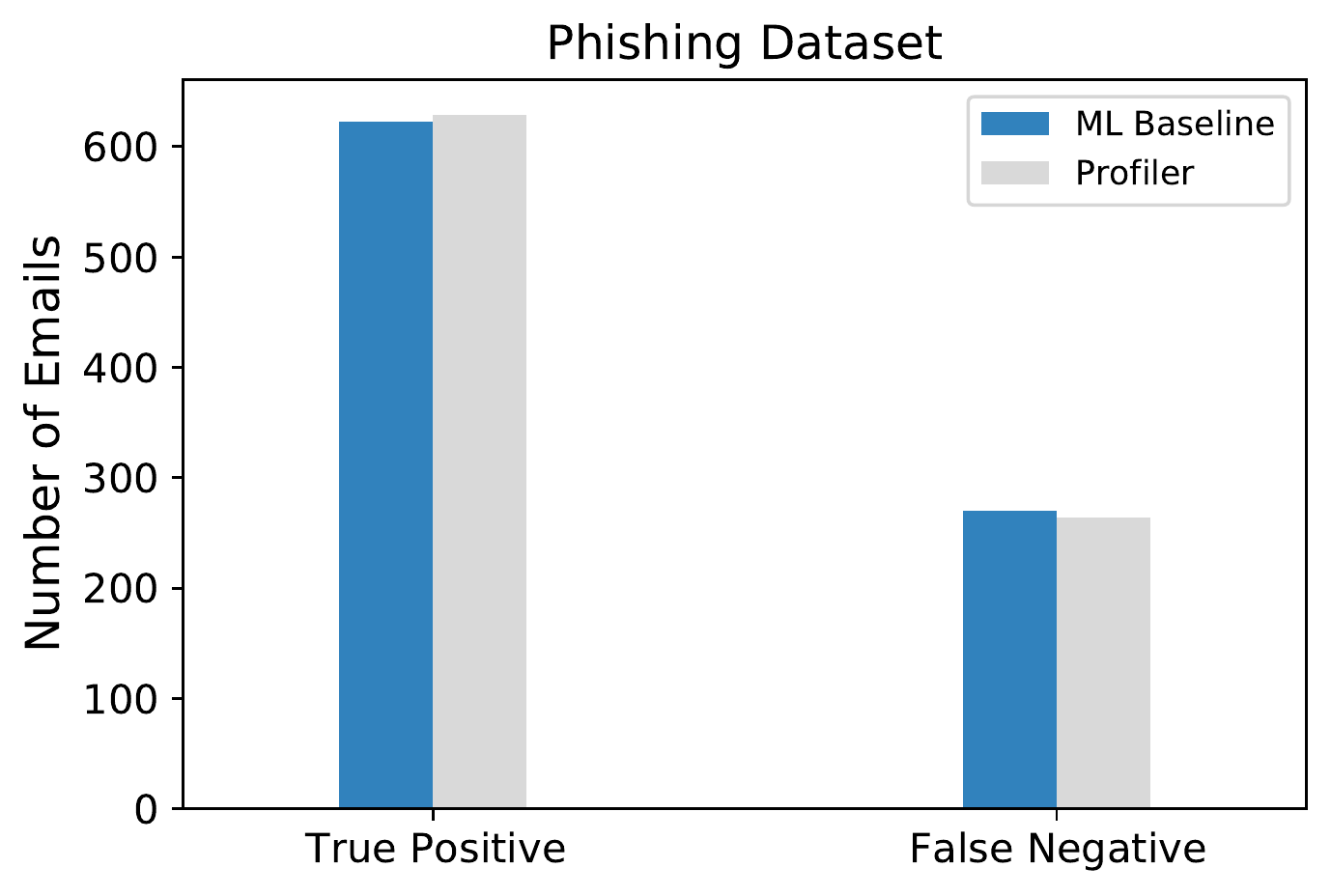}
\caption{Comparing Machine Learning Baseline to Profiler Classifications (n = 893).}
\label{fig:phishingresults}
\end{figure}

Similarly, the Profiler labelled 70.4\% of this data set correctly: 629 labelled true positive and 264 false positive labels. We can see these results visually in Figure \ref{fig:phishingresults} with the true positive column on the left and the false negative column on the right. 

The machine learning ensemble and the Profiler achieved indistinguishable results on this data set. However, THEMIS, one of the algorithms used in the machine learning ensemble, trained on a significant portion of this data set which may have inflated the accuracy of the ensemble. We hypothesis that the machine learning baseline result has a higher true positive than would have otherwise been achieved if THEMIS had not trained on this data. 

\subsection{File Sharing Phishing Data Set Results}
With today's online world, file sharing services such as Google Drive, Dropbox, Microsoft One Drive and Sharepoint are booming. They allow us to share files with ease between different people and to collaborate on and sync work to the cloud. These services often communicate any sharing of files, folders or documents with their user base over email. As such, attackers have taken advantage of this and launched a new type of phishing attack: file sharing phishing. This attack is successful because the emails contain legitimate links to file sharing services which cannot be detected by existing phishing detection solutions. Machine learning approaches which look at URLs are ineffective and are unable to recognise the threat.

This form of phishing is relatively new and to the best of our knowledge, there are no targeted solutions towards it yet. This could be due to the lack of file sharing phishing data, or to the sheer amount of resources that it takes to train a new algorithm. 

The file sharing phishing data set contains 3292 file sharing phishing emails which were unseen by both the machine learning models in the ensemble and us during the development of the Profiler. The machine learning ensemble labeled 55.2\% correctly: 1817 emails were labelled true positive and 1475 were false negative labels.

\begin{figure}[t]
\centering  
\includegraphics[width=0.8\linewidth]{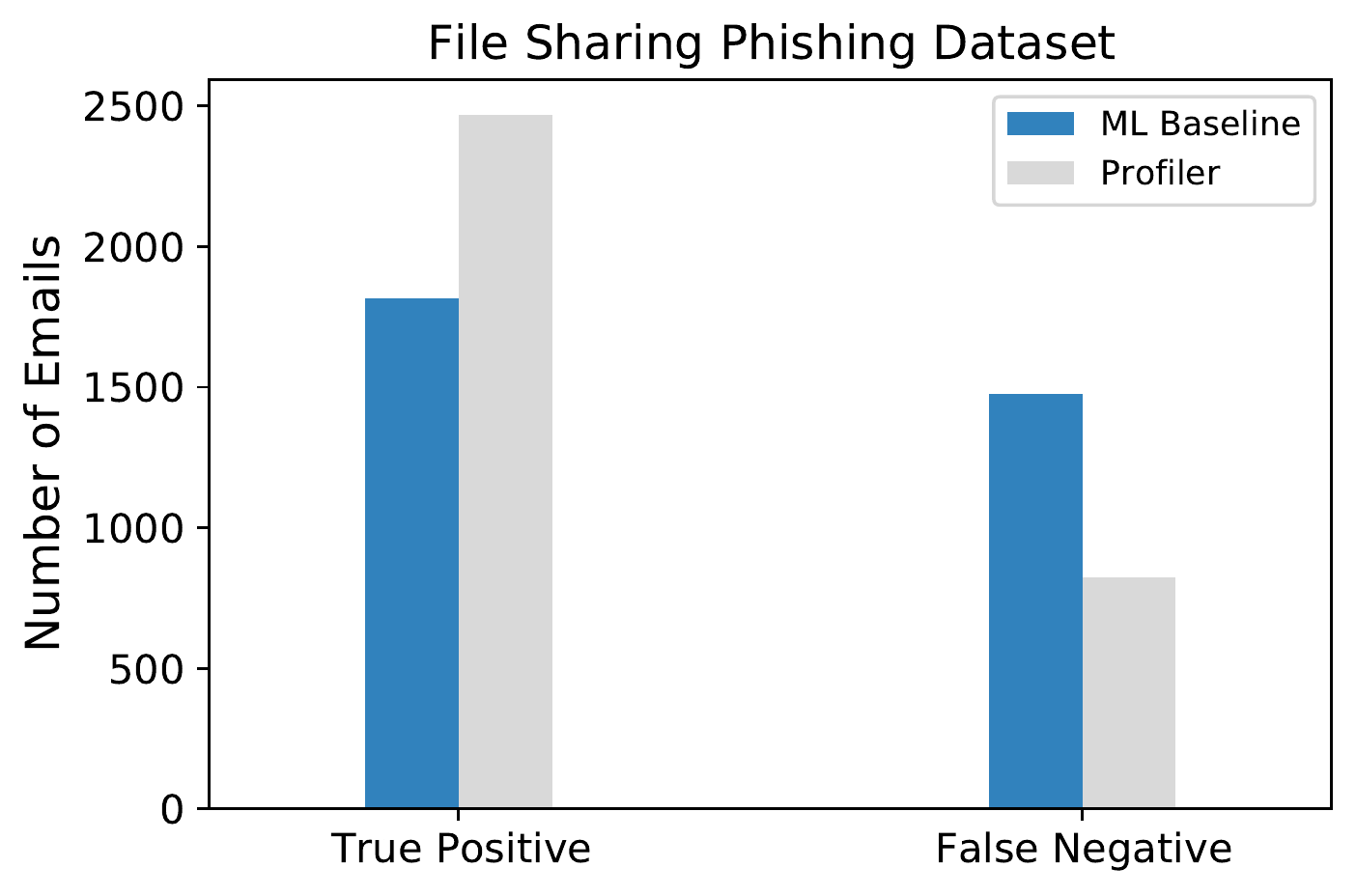}
\caption{Comparing Machine Learning Baseline to Profiler Classifications (n = 3292).}
\label{fig:filesharingphishingresults}
\end{figure}

The Profiler labelled 75\% of this data set correctly: 2469 labelled true positive and 823 false positive labels. We can see these results in Figure \ref{fig:filesharingphishingresults} with the true positive column on the left and the false negative column on the right.

The machine learning models within our ensemble had a reported accuracy of more than 90\%, yet our experiments had high volumes of false negative classifications. We hypothesis that this is due to the fact that file sharing phishing is a new form of phishing and the three algorithms, THEMIS \cite{fang2019themis}, URLNet \cite{le2018urlnet}, and HTMLPhish \cite{opara2019htmlphish}, are suffering from concept drift - they are unable to recognise the potential threat of the email as they have never trained to recognise such emails. 

With the Profiler, we see that there is a 20\% improvement in true positive classifications on this data set. We suggest that with our horizontal approach we can detect new forms of email phishing and not suffer from concept drift as vertical approaches do.

\section{Discussion}
\label{section:discussion}

In this section we discuss how the Profiler can be used in conjunction with machine learning models in order to leverage the benefits of the horizontal approach over the vertical approach. We also discuss the security, performance, and adaptability of the Profiler and present future work. 

\subsection{Implication of Results}
The novelty of the Profiler lies in its horizontal approach to email phishing detection - heuristically profiling an email on its features to determine the risk of it being a phishing attack. Given the sensitive nature of emails and various privacy acts, access to large email data sets on which to train on is not readily available. However, the Profiler does not require training and thus, a smaller, focused email data set is all that is needed. Further, by accounting for multiple features of an email, the Profiler is able to avoid the impact of concept drift which occurs as a result of the constant evolution of phishing attacks. 

The Profiler framework's numerical output makes it easy to integrate with existing machine learning models to help improve email phishing detection. It can reliably identify real time emails which may be false positive or false negative classifications such that they can be passed through selected machine learning models for further triaging. Its three models can also be improved upon, or other models may be added to help identify new forms of email phishing. 

Further, the Profiler can be used in the training stage of machine learning model development to help with labelling of large email data sets. Passing the data through the Profiler, each email can get it's own risk assessment score. Then, according to a pre-determined threshold, the Profiler can label emails as either legitimate or phishing, reducing the time taken to re-train machine learning models and helping improve real time accuracy of the models. 

\subsection{Privacy and Security}

Gaining access and storing large data sets of emails can be challenging due to their sensitive nature and the strictness of privacy acts such as the GDPR and the Australian Privacy Act. Email collection for a machine learning approach falls under these privacy acts, with penalties for infringement. In order to use email data, there must be, first, strict permission from the user, and strict protocols must be followed. Personal data cannot be reused indiscriminately, all decision-making must be explainable and there must be a timeline for deletion of data \cite{gdprantipatterns}. All this makes the finding and storing of data difficult. 

Further, storage of data creates opportunities for security breaches. By holding on to large email data sets for training, the security of the information within the email is exposed. Attackers could target organisations that are using these large data sets and gain access to them, resulting in the loss of confidential information. 

The Profiler has an advantage in that there is no need to store email data after it has been evaluated, and there is no pre-training of the model which would require large data sets to be stored and accessed. As such, there does not need to be considerations on the security and protection of large data sets. 

\subsection{Performance}
A rule-based system can execute decisions much faster than a machine learning approach, and without prior training. In an area such as email phishing detection, where data sets can contain 10,000+ emails, speed of execution becomes a vital consideration. 

Taking one of the algorithms used in the machine learning ensemble as an example, HTMLPhish must go to the URL given as its input, load its web page and then evaluate the HTML document, taking up to 1.4 seconds \cite{opara2019htmlphish}. Performing this operation for each URL in an email, over many email inputs can take up a lot of time and computing power.

In contrast, the Profiler performs in linear time, dependent on the length of the body of the email and can evaluate an email in approximately 0.3 seconds. The Profiler's faster execution speeds can be crucial when evaluating large volumes of incoming emails.

\subsection{Adaptability}
On average, 40\% of companies take more than a month to deploy a machine learning model into production and 78\% of machine learning projects which involve training stalls before deployment \cite{newstackio}. This goes to show how difficult it can be to train and develop a new machine learning model.

Given that forms of email phishing are continuously evolving, this makes the creation of new machine learning models which can recognise new forms of email phishing or re-training of existing models a challenging task. A major benefit of the Profiler is its modular design and adaptability. 

\subsubsection{Modular Design} 
Currently, the Profiler considers three models however, this could be easily extended to more or different models. For example, a model which considers the BCC'd and CC'd receivers of an email could be added, or a machine learning model which does sentiment analysis on the language could replace or work in conjunction with the existing cognitive assessment model. Another benefit of the Profiler, is that the development of new models can be done in parallel, and without reliance on other models. There is undeniable flexibility involved in this solution. 

\subsubsection{Sliding Threshold}
Recall that the output of the Profiler is a numerical output greater than or equal to 0. In this work, we used a threshold of 0.5 - anything below 0.5 was labelled legitimate and anything greater than or above was phishing. However, the threshold does not have to be 0.5 and can be easily adjusted for different use cases. This creates great flexibility in our solution. 

For example, a lower threshold of 0.2 could be used to reduce the false positive classifications. This situation can also be reversed by using a higher threshold, for example, of 5. In this manner, we would capture mainly phishing emails and reduce our false negatives. 

Another benefit of a numerical output is that there can be multiple thresholds on the scale. For example, consider two thresholds: a lower threshold of 0.3 where anything below it is classified as legitimate, and another, a higher threshold at 0.9 which dictates where the phishing email classification begins. In this way, we can either ignore the emails which score between these points (as they have more chance of being false positive or false negative classifications), or we can pass them through to machine learning models for further investigation. 

\subsection{Underlying Heuristic Approach vs Machine Learning}
In this work, we implemented our three models using heuristic methods to demonstrate the benefits of a horizontal approach. Our three models were capable of performing at or above the level of a machine learning majority-voting ensemble. 

However, the purpose of this work is not to discount machine learning as an email phishing detection solution. Rather, we have identified challenges which machine learning models face, such as lack of data and a constantly evolving application domain, and propose a more horizontal approach to email phishing detection. 
If appropriate data sets were found on which to train machine learning models, then we could implement them into our Profiler. The idea is not to replace machine learning, rather to look at more features of an email and not focus on training vertically - which is the cause of concept drift. 

\subsection{Future Work}

This paper is the first step to approaching email phishing detection in a horizontal fashion. Though our results demonstrate its benefits and improvement in accuracy, there is still an opportunity for future work, specifically in the development of more models and exploring integration with machine learning solutions. 

\subsubsection{More Models}
Given its modular design, more models can be easily added to the Profiler. These can be models which focus specifically on new and evolving forms of phishing such as a targeted model for identifying file sharing phishing, or models which look at other features of emails. Already, we discussed BCC'd and CC'd receivers and we could further explore other important characteristics on which to profile an email such as time received, familiarity with sender, sentiment analysis, etc.  

\subsubsection{Integration with Machine Learning Methods}

We see two possibilities for integrating the Profiler with machine learning solutions.

First, the Profiler can be used in conjunction with machine learning models to help improve email phishing detection. Using the Profiler's risk assessment output per email, we could identify those which are more likely to be false positive and false negative classifications and pass them through to targeted machine learning solutions.

In this manner, we can filter out emails which we have high confidence to be legitimate and phishing emails, and concentrate all resources on those where there is a possibility for misclassification. Not only could this improve the accuracy of email phishing detection, but also it would improve the performance, as less computing power would be required. 

Second, the Profiler can be used to help label large data sets of emails to improve the speed of training machine learning models. Large data sets of emails that are used by research organisations need to be manually marked as legitimate or phishing emails, taking valuable time and resources away from the development of the actual machine learning model. 

As such, the Profiler could be utilised to automatically label these data sets. Each user of the Profiler can determine their own thresholding which would suit their use case. This would enable machine learning approaches to adapt more rapidly to constantly evolving phishing attacks. Further, the Profiler score can be used as an additional feature on which to train machine learning models.

These two applications can be investigated in future works and evaluations to further demonstrate the Profiler's capabilities. 

\section{Conclusion}

In this paper, we explore a horizontal approach to email phishing detection. Our analysis covered three data sets totalling to 14,000 emails and examined the implication of concept drift and lack of data on the effectiveness of the vertical approach to email phishing detection that most machine learning algorithms take. We show that there is a gap between the reported accuracy from literature and the algorithms' actual effectiveness in the real world. We argue that the Profiler sees better accuracy and performance measures than existing solutions and, as such, propose it as a standalone solution or adjoining tool to improve machine learning email phishing detection. 

\bibliographystyle{unsrt}
\bibliography{references}

\end{document}